%% file: colm2024_conference.tex
\documentclass{article} %
\usepackage{colm2024_conference}

\usepackage{booktabs}
\usepackage{graphicx}
\usepackage{enumitem}
\usepackage{wrapfig}
\usepackage{algorithm}
\usepackage{algpseudocode}

\usepackage{microtype}
\usepackage{amsmath}
\usepackage{amsfonts}
\usepackage{colortbl}
\usepackage[utf8]{inputenc}
\usepackage[T1]{fontenc}
\definecolor{lightgray}{rgb}{0.9,0.9,0.9}
\usepackage{caption}
\usepackage{subcaption}
\usepackage{xcolor}
\usepackage{setspace}
\usepackage{url}
\usepackage{multirow}
\usepackage{colortbl}
\usepackage{tabularx}
\usepackage{blindtext}
\usepackage{pgfplots}
\pgfplotsset{compat=1.18} 
\usepackage{tikz}
\usetikzlibrary{er,positioning,bayesnet}
\usepackage{makecell}
\usepackage{tipa}
\usepackage{siunitx}
\usepackage{nicefrac}
\usepackage{tocloft}
\usepackage{listings}
\usepackage[raster,skins]{tcolorbox} %
\usepackage{xltabular}
\usepackage{adjustbox}
\usepackage{xurl}
\usepackage{xcolor}
\input{math_commands.tex}

\title{Qwen3-TTS Technical Report}

\author{
\bf Qwen Team
}

\newcommand{\method}{Qwen3-TTS\xspace}

\begin{document}

\maketitle

\begin{abstract}
In this report, we present the Qwen3-TTS series, a family of advanced multilingual, controllable, robust, and streaming text-to-speech models. Qwen3-TTS supports state-of-the-art 3-second voice cloning and description-based control, allowing both the creation of entirely novel voices and fine-grained manipulation over the output speech. Trained on over 5 million hours of speech data spanning 10 languages, Qwen3-TTS adopts a dual-track LM architecture for real-time synthesis, coupled with
two speech tokenizers: 
1) \textit{Qwen-TTS-Tokenizer-25Hz} is a single-codebook codec emphasizing semantic content, which offers seamlessly integration with Qwen-Audio and enables streaming waveform reconstruction via a block-wise DiT.
2) \textit{Qwen-TTS-Tokenizer-12Hz} achieves extreme bitrate reduction and ultra-low-latency streaming, enabling immediate first-packet emission ($97\,\mathrm{ms}$) through its 12.5 Hz, 16-layer multi-codebook design and a lightweight causal ConvNet.
Extensive experiments indicate state-of-the-art performance across diverse objective and subjective benchmark (e.g., TTS multilingual test set, InstructTTSEval, and our long speech test set). To facilitate community research and development, we release both tokenizers and models under the Apache 2.0 license.

\end{abstract}

\input{content/intro.tex}
\input{content/tokenizer.tex}

\input{content/method}
\input{content/experiments.tex}

\input{content/conclusion.tex}

\input{content/authors.tex}

\bibliography{biblio}
\bibliographystyle{colm2024_conference}

\end{document}

%% file: math_commands.tex
\usepackage{amsmath,amsfonts,bm}

\def\eqref#1{equation~\ref{#1}}

\def\1{\bm{1}}

\DeclareMathAlphabet{\mathsfit}{\encodingdefault}{\sfdefault}{m}{sl}
\SetMathAlphabet{\mathsfit}{bold}{\encodingdefault}{\sfdefault}{bx}{n}



%% file: content/intro.tex
\section{Introduction}
\label{sec:intro}
\begin{figure}[h]
    \centering
    \includegraphics[width=0.98\textwidth]{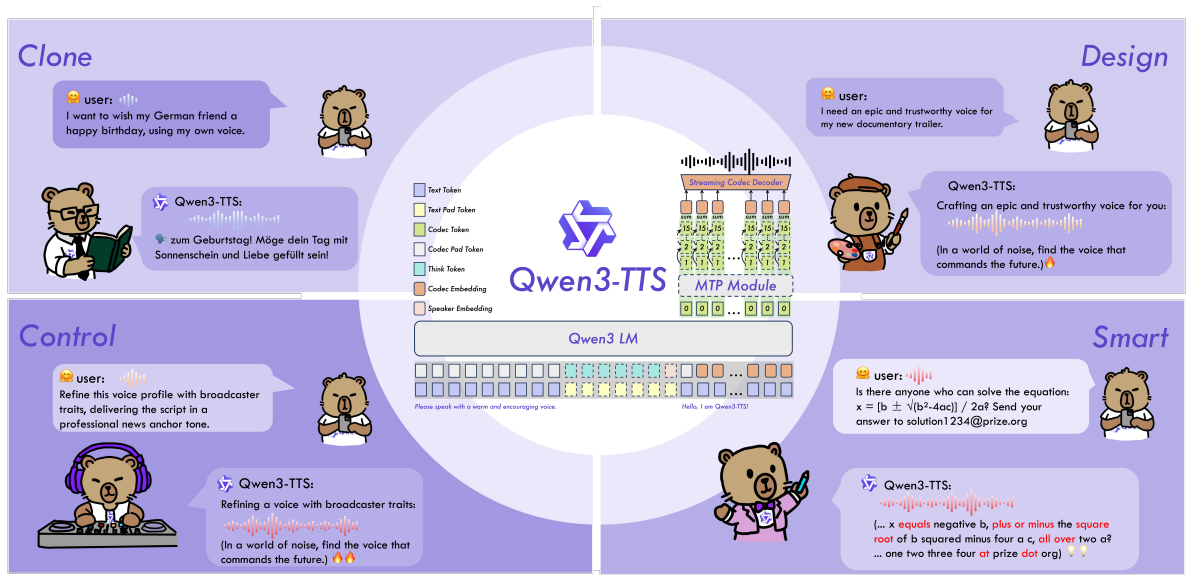}
    \caption{\method is a multilingual, controllable, robust, and streaming text-to-speech model. Based on these features, \method supports a wide range of tasks, including but not limited to cloning, creating and controlling voice, and easily handling various complex texts.}
    \label{fig:ovewview_arch}
\end{figure}

Stable, controllable, and human-like speech synthesis is widely viewed as a key capability on the path to AGI. Modern neural text-to-speech (TTS) models, trained on large-scale datasets, already deliver exceptional capability to generate high-quality speech from a few seconds of reference audio~\citep{valle,naturalspeech2,naturalspeech3,uniaudio,e2tts,f5tts,cosyvoice,sparktts,maskgct,llasa}. Among them, discrete speech tokenization~\citep{Encodec,SoundStream,dac} combined with autoregressive language modeling of discrete units has gained traction, offering improved stability while preserving high naturalness and human-likeness~\citep{cosyvoice3,fish-speech-v1.4,moshi,qwen2.5omni,minimax-speech}. Conditioning on vocal features or text instructions facilitates finer-grained control over prosody and style, resulting in outputs of greater richness and diversity~\citep{cosyvoice2,parlertts,voxinstruct,controlspeech,prompttts2}. These breakthroughs are paving the way for diverse applications in fields such as virtual assistants and automated content creation.

\begin{table}[t!]
\footnotesize
\centering
\caption{\textbf{Overview of the \method model family.}}
\begin{tabular}{@{}lcccc@{}}
\toprule
\textbf{Model Name}                 & \textbf{Streaming}    & \textbf{Multilinguality} & \textbf{Voice Clone}  & \textbf{Instruction Following} \\ \midrule
\method-12Hz-1.7B-Base     & $\checkmark$ & $\checkmark$  & $\checkmark$ &                           \\
\method-12Hz-1.7B-VoiceDesign & $\checkmark$ & $\checkmark$  &              & $\checkmark$              \\
\method-12Hz-1.7B-CustomVoice      & $\checkmark$ & $\checkmark$  &              & $\checkmark$              \\
\method-12Hz-0.6B-Base     & $\checkmark$ & $\checkmark$  & $\checkmark$ &                           \\
\method-12Hz-0.6B-CustomVoice      & $\checkmark$ & $\checkmark$  &              &                         \\
\method-25Hz-1.7B-Base     & $\checkmark$ & $\checkmark$  & $\checkmark$ &                           \\
\method-25Hz-1.7B-VoiceEditing & $\checkmark$ & $\checkmark$  & $\checkmark$ & $\checkmark$              \\
\method-25Hz-1.7B-CustomVoice      & $\checkmark$ & $\checkmark$  &              & $\checkmark$              \\
\method-25Hz-0.6B-Base     & $\checkmark$ & $\checkmark$  & $\checkmark$ &                           \\
\method-25Hz-0.6B-CustomVoice      & $\checkmark$ & $\checkmark$  &              &                           \\ \bottomrule
\end{tabular}
\end{table}


In this report, we take a step toward stable, controllable, and human-like speech synthesis and introduce \method, the first text-to-speech model in the Qwen series. \method exhibits the following properties: 
1)~\textbf{Controllability}: \method allows users to create new voices or manipulate fine-grained attributes of generated speech via natural language descriptions, while also supporting the stable generation of any content using the created voice. 2)~\textbf{Voice Cloning and Predefined Voice Profiles}: \method supports 3-second voice cloning and generation using a set of x curated, high-quality preset voices.
3)~\textbf{Naturalness}: Beyond achieving robust synthesis, \method excels in generating highly natural and expressive speech. Our 1.7B model, in particular, delivers state-of-the-art, human-like quality, demonstrating our approach successfully maximizes perceptual quality without overfitting to ASR-related metrics.
4)~\textbf{Multilinguality}: The model is trained across more than 10 languages and supports speaker-consistent multilingual generation.
5)~\textbf{Streaming}: Designed for streaming text input and streaming audio output, it achieves a first-packet latency as low as 97 ms (0.6B variant) and 101 ms (1.7B variant).




Beyond the aforementioned aspects, and from a broader perspective of practical application, it is crucial for our model to integrate seamlessly with Large Language Models (LLMs) and achieve extremely low first-packet latency. To this end, we use discrete speech representations as the cornerstone of our architecture and introduce two tokenizers in the \method family: 1) \textit{Qwen-TTS-Tokenizer-25Hz} employs a 25 Hz single-codebook representation with waveform reconstruction via block-wise flow matching to enable streaming synthesis~\citep{qwen2.5omni}. Empirically, we find that semantic tokenizers lack expressive power, whereas purely acoustic tokenizers inject excessive low-level detail that complicates LLM-based modeling and leads to long-horizon error accumulation. To balance these factors, Qwen-TTS-Tokenizer-25Hz integrates semantic and acoustic cues, leveraging the Qwen2-Audio encoder for both expressivity and tractability. Although it supports streaming with a block-wise diffusion decoder, we found that its single-codebook design limits suitability for ultra-low-latency applications and general speech synthesis. Therefore, we develop 2) \textit{Qwen-TTS-Tokenizer-12Hz}, which adopts a 12.5 Hz multi-codebook scheme inspired by~\citet{zhang2023speechtokenizer}. Its first codebook layer encodes semantic content, while the subsequent layers capture acoustic details. The increased capacity permits waveform reconstruction using only a lightweight causal ConvNet, eliminating the need for speaker vector extraction or complex diffusion models~\citep{cosyvoice2,minimax-speech}. To further support ultra–low-latency streaming, we designed a dual-track autoregressive architecture for streaming text input and audio output. This architecture incorporates a Multi-Token Prediction (MTP) module to effectively model the multi-codebook sequence, which enables immediate speech decoding from the first codec frame.




Trained on over 5 million hours of speech data, \method achieves impressive performance across diverse benchmarks. Specifically, it establishes a new state-of-the-art in zero-shot voice cloning, achieving the lowest Word Error Rate (WER) on the Seed-TTS benchmark while delivering superior speaker similarity across all 10 evaluated languages compared to commercial baselines like MiniMax and ElevenLabs. In cross-lingual scenarios, \method demonstrates exceptional adaptability, reducing error rates by significant margins in challenging pairs such as Chinese-to-Korean. Regarding controllability, \method excels in following complex natural language instructions for voice design and control, outperforming GPT-4o-mini-tts in target speaker manipulation. Furthermore, the model exhibits remarkable stability in long-form generation, capable of synthesizing over 10 minutes of natural and fluent speech. To facilitate community research and development, we release the complete family of \method models and tokenizers.

%% file: content/tokenizer.tex
\section{Qwen-TTS-Tokenizer}

\subsection{Qwen-TTS-Tokenizer-25Hz}
\begin{figure}
        \centering
        \includegraphics[width=\textwidth]{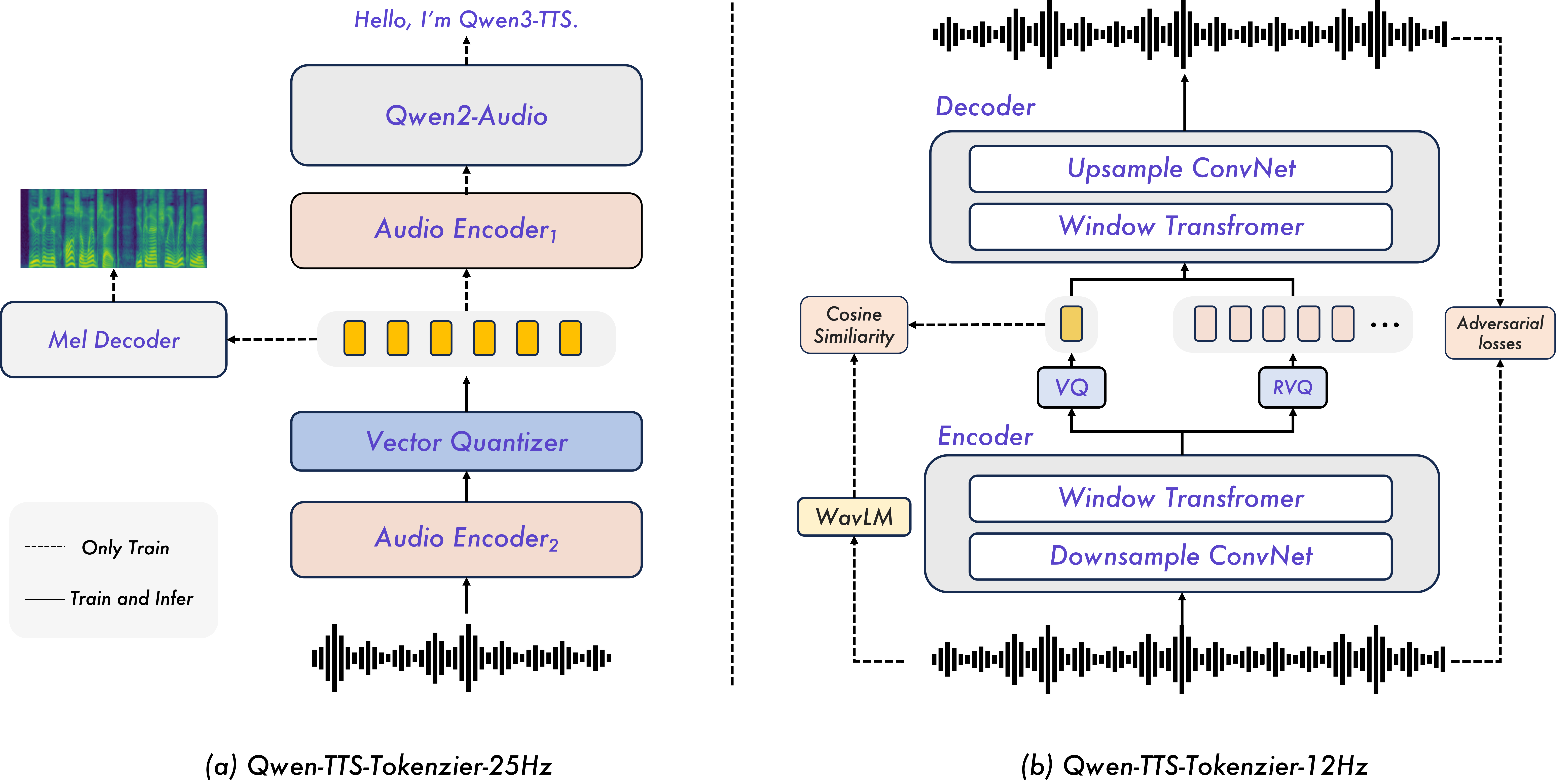}
        \label{fig:tokenizer}
    \caption{Overview of Qwen-TTS tokenizers.}
    \label{fig:overview_arch}
\end{figure}
\paragraph{Tokenizer}
Qwen-TTS-Tokenizer-25 Hz is a 25 Hz single-codebook tokenizer built upon Qwen2-Audio through a two-stage training framework. In the first stage (Stage 1), we continue pretraining Qwen2-Audio on an automatic speech recognition (ASR) task, augmenting the audio encoder with an additional resampling layer and a vector quantization (VQ) layer inserted at an intermediate position. In the second stage (Stage 2), we fine-tune the entire model by incorporating a convolution-based mel-spectrogram decoder, which reconstructs mel-spectrograms from the audio tokens. This reconstruction objective explicitly injects essential acoustic information into the learned audio token representations.
\paragraph{Streaming Detokenizer}
To enable streaming audio generation, particularly for long sequences, we propose a sliding-window block attention mechanism that restricts each token to a limited context. Specifically, we use a Diffusion Transformer (DiT) trained with Flow Matching~\citep{flowmatching}. The input code sequence is first mapped to a mel-spectrogram via Flow Matching, after which a modified BigVGAN~\citep{sangbigvgan} reconstructs the waveform from the generated mel-spectrogram. 

To support streaming decoding, we group adjacent codes into fixed-length blocks and construct the corresponding attention mask~\citep{streamflow}. The DiT’s receptive field is restricted to 4 blocks—the current block, a 3-block lookback, and a 1-block lookahead. During decoding, we generate mel-spectrograms in chunks with Flow Matching, ensuring that each code chunk has access to the required context blocks. This design improves streaming quality by preserving necessary context. We apply the same chunked procedure to BigVGAN, whose receptive field is fixed, to support streaming waveform synthesis.

\subsection{Qwen-TTS-Tokenizer-12Hz}

Qwen-TTS-Tokenizer-12Hz is a 12.5 Hz multi-codebook tokenizer with jointly optimized semantic and acoustic streams. Building on the semantic–acoustic disentangled quantization strategy of the Mimi architecture~\citep{moshi}, speech is decomposed into two discrete code sequences: a semantic codebook capturing high-level semantic content and an acoustic codebook modeling acoustic detail, prosody, and others. Training adopts a GAN-based framework in which the generator operates directly on raw waveforms to extract and quantize both representations, while the discriminator improves the naturalness and fidelity of reconstructed speech. A multi-scale mel-spectrogram reconstruction loss further enforces time–frequency consistency. For the semantic path, WavLM~\citep{wavlm} serves as a teacher to guide the first semantic codebook layer toward semantically aligned features. The acoustic path employs a 15-layer residual vector quantization (RVQ) module that progressively refines details not captured by the semantic codebook. To enable streaming, we use fully causal feature encoders and decoders: the encoder processes frames sequentially and emits semantic and acoustic tokens at 12.5 Hz without look-ahead, and the decoder reconstructs audio incrementally from these tokens. This end-to-end causal design supports streaming inference with low latency, making the tokenizer suitable for real-time online services.


%% file: content/method.tex
\section{Method}
\begin{figure}[tbh]
    \centering
    \includegraphics[width=0.8\textwidth]{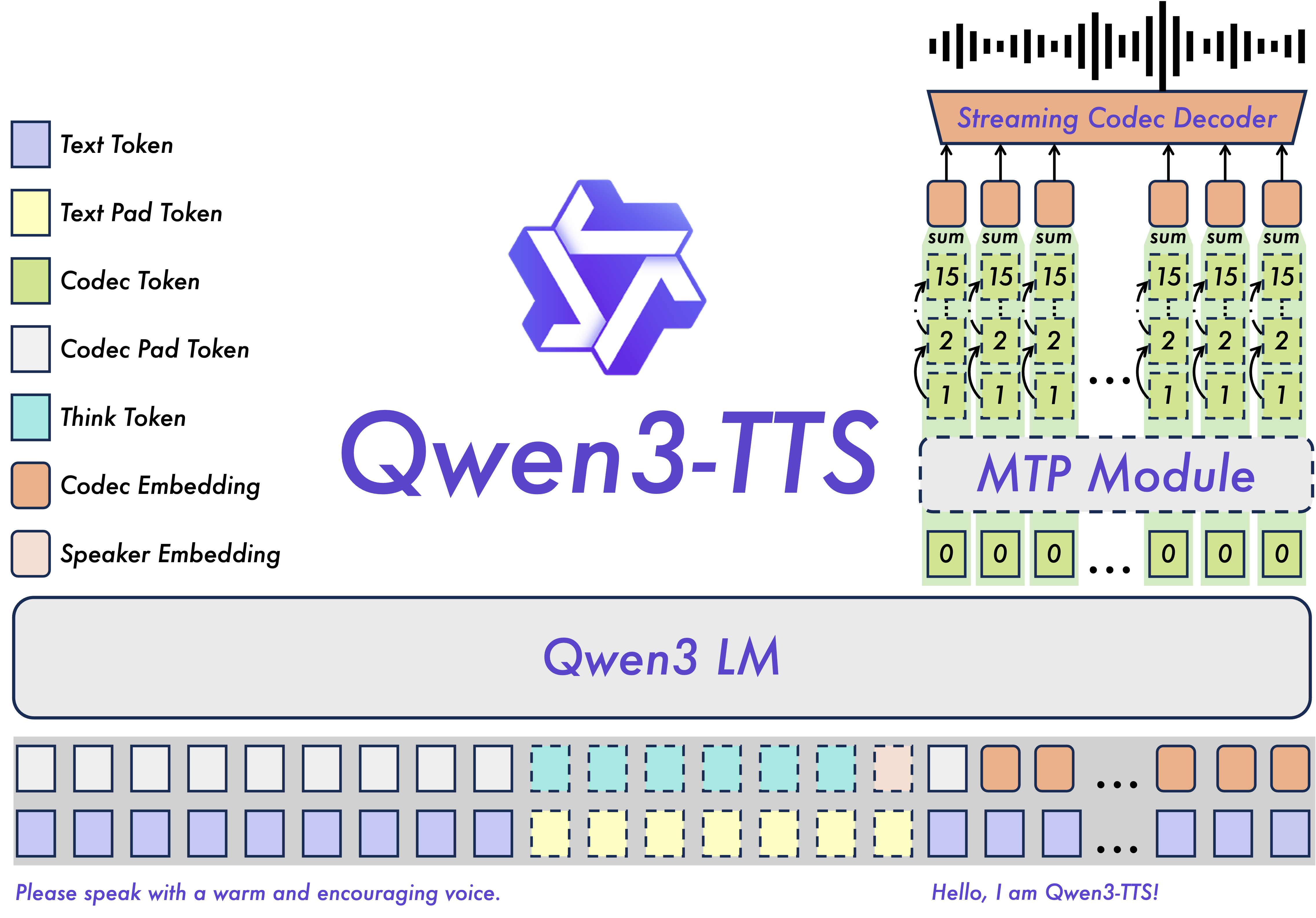}
    \caption{The overview of \method. Dashed lines represent optional.}
    \label{fig:ovewview_arch}
\end{figure}

\subsection{Architectures}
\method leverages the Qwen3 LM family to achieve high concurrency and low-latency inference. Text is processed using the standard Qwen tokenizer, while speech is encoded using the Qwen-TTS-Tokenizer. To maintain precise identity control, we jointly train a learnable speaker encoder with the backbone. For real-time synthesis, \method employs a dual-track representation by concatenating textual and acoustic tokens along the channel axis. Upon receiving a textual token, the model immediately predicts the corresponding acoustic tokens, which are then converted into waveforms by the Code2Wav module.

\paragraph{Qwen3-TTS-25Hz}
\method-25Hz uses Qwen-TTS-Tokenizer-25Hz to extract single-level speech tokens. The backbone integrates text features with preceding speech tokens and predicts the current speech token through a linear head. The resulting sequence is then processed by a chunk-wise DiT module for high-fidelity waveform reconstruction.

\paragraph{Qwen3-TTS-12Hz}
Architecturally, \method-12Hz differs from \method-25Hz by operating on RVQ tokens from Qwen-TTS-Tokenizer-12Hz. It adopts a hierarchical prediction scheme: the backbone ingests aggregated codebook features to predict the zeroth codebook, and an MTP (Multi-Token Prediction) module then generates all residual codebooks. This strategy captures intricate acoustic details, significantly enhancing vocal consistency and expressivity, while minimizing latency through single-frame instant generation.

\subsection{Training}
The training process consists of pre-training and post-training. All data is formatted in ChatML to standardize inputs and support controllable speech generation.

The pre-training of \method is structured into three stages:
\begin{enumerate}[label=(\arabic*)]
    \item \textbf{General Stage (S1)}: During the initial pre-training phase, we leverage over 5 million hours of multilingual speech data to train \method. This stage establishes a monotonic mapping from multilingual text representations to speech and builds general capabilities for \method.
    \item \textbf{High-Quality Stage (S2)}: We stratify data quality with a dedicated pipeline and perform continual pre-training (CPT) with high-quality data. This stage alleviates hallucinations caused by noisy data in the initial stage and significantly improves the quality of generated speech.
    \item \textbf{Long-Context Stage (S3)}: In the final pre-training phase, we increase the maximum token length from 8{,}192 to 32{,}768 and upsample long speech in the training data. Experimental results indicate that these adjustments enhance the model's ability to process extended and complex inputs and to generate contextually appropriate speech responses.
\end{enumerate}

The post-training phase comprises three stages, enabling \method to generate human-like speech and remain stable across tasks. In the first stage, we introduce Direct Preference Optimization (DPO)~\citep{rafailov2024direct} to align model outputs with human preferences. Specifically, we construct preference pairs for multilingual speech samples based on human feedback and then perform DPO on \method. In the second stage, we employ rule-based rewards and leverage GSPO to comprehensively enhance the model's capabilities and stability across tasks. Finally, we introduce lightweight speaker fine-tuning on the base model, enabling \method to adopt specific voices while further improving the naturalness, expressiveness, and controllability of its speech responses.

\subsection{Features}
\method supports streaming voice cloning, voice design, and fine-grained control. To achieve this, we prepend user-provided instructions containing fine-grained control signals to the input sequences.

For voice cloning, \method clones a target voice from (i) reference speech via a speaker embedding, enabling real-time cloning, or (ii) a text--speech pair via in-context learning, which better preserves prosody.
For voice design, built upon the Qwen3 text model foundation, \method inherits robust text comprehension capabilities. Additionally, we introduce a probabilistically activated \emph{thinking pattern} during training to improve instruction following, especially for complex descriptions. Furthermore, based on this strong instruction-following capability, \method controls predefined voices with desired styles.

\subsection{Efficiency}

Low first-packet latency and stable streaming under concurrency are jointly determined by (i) the language model (LM) time to first token group for the first speech packet, and (ii) the tokenizer decoding pipeline that converts generated tokens into waveforms. As shown in Table~\ref{tab:qwen3tts_efficiency}, we evaluate Qwen3-TTS with different LM sizes and tokenizer variants under various concurrency levels. All reported numbers are end-to-end measured latencies, and steady-state costs are measured per speech packet during streaming generation. Specifically, latency is measured on our internal vLLM engine (vLLM V0 backend) on a single typical computational resource with optimizations applied via \textit{torch.compile} and CUDA Graph acceleration to the decoding stage of the tokenizer. Reported First-Packet Latency is the sum of LM time-to-first packet tokens (TTFP) and tokenizer decode time for per-packet (TPP). LM time for per-packet (TPP) is the steady-state LM time to produce one packet's tokens during streaming generation.

\begin{table}[!h]
\centering
\caption{\textbf{Streaming efficiency of Qwen3-TTS with different tokenizers under varying concurrency.}}
\vspace{-1mm}
\footnotesize
\setlength{\tabcolsep}{3pt}
\begin{tabular}{lcccccc}
\toprule
\textbf{Model} & \textbf{Concurrency} &
\textbf{LM TTFP} &
\textbf{Tokenizer Decode TPP} &
\textbf{First-Packet Latency} &
\textbf{LM TPP} &
\textbf{RTF} \\
\midrule
Qwen3-TTS-25Hz-1.7B & 1 & 125~ms & 25~ms  & 150~ms & 56~ms & 0.253 \\
Qwen3-TTS-25Hz-1.7B & 3 & 222~ms & 62~ms  & 284~ms & 64~ms & 0.394 \\
Qwen3-TTS-25Hz-1.7B & 6 & 376~ms & 147~ms & 523~ms & 85~ms & 0.725 \\
\midrule
Qwen3-TTS-25Hz-0.6B & 1 & 113~ms & 25~ms  & 138~ms & 50~ms & 0.234 \\
Qwen3-TTS-25Hz-0.6B & 3 & 198~ms & 62~ms  & 260~ms & 59~ms & 0.378 \\
Qwen3-TTS-25Hz-0.6B & 6 & 334~ms & 147~ms & 481~ms & 80~ms & 0.709 \\
\midrule
Qwen3-TTS-12Hz-1.7B & 1 & 97~ms  & 4~ms & 101~ms & 21~ms & 0.313 \\
Qwen3-TTS-12Hz-1.7B & 3 & 190~ms & 5~ms & 195~ms & 24~ms & 0.363 \\
Qwen3-TTS-12Hz-1.7B & 6 & 328~ms & 5~ms & 333~ms & 32~ms & 0.463 \\
\midrule
Qwen3-TTS-12Hz-0.6B & 1 & 93~ms  & 4~ms & 97~ms  & 19~ms & 0.288 \\
Qwen3-TTS-12Hz-0.6B & 3 & 174~ms & 5~ms & 179~ms & 22~ms & 0.338 \\
Qwen3-TTS-12Hz-0.6B & 6 & 294~ms & 5~ms & 299~ms & 30~ms & 0.434 \\
\bottomrule
\end{tabular}
\label{tab:qwen3tts_efficiency}
\end{table}

Qwen-TTS-Tokenizer-25Hz performs code-to-waveform synthesis through chunk-wise inference. Due to the look-ahead requirement in the DiT module, waveform synthesis for the first chunk cannot start until sufficient future tokens are available. With a chunk size of 8 set in Qwen3-TTS, the model must wait for the LM to generate 16 tokens before DiT can produce the first 8-token mel chunk. Under the 25 Hz token rate (40 ms per token), this corresponds to 320 ms of mel content per packet. In addition, the BigVGAN vocoder introduces an extra right-context look-ahead (130 ms). Therefore, for Tokenizer-25Hz, the first packet ultimately contains about 190 ms of audio, and the LM must generate 16 tokens before synthesis can start. During steady-state streaming generation, every time the LM generates 8 tokens, DiT and BigVGAN can synthesize a 320 ms audio packet. The first-packet latency and RTF reported in our table are computed based on the above setup.

Qwen-TTS-Tokenizer-12Hz uses a pure left-context streaming codec decoder, enabling waveform emission immediately after the required tokens are available, without waiting for future context. With the 12.5 Hz token rate, each token corresponds to 80 ms of audio, so one token can be decoded into audio directly in principle. To avoid excessive scheduling overhead caused by very small packets, we define one speech packet as 4 tokens, which means 320 ms of speech per packet. This design significantly reduces decoding time and yields lower first-packet latency, while maintaining low RTF under higher concurrency due to the lightweight and batch-friendly codec decoder.

%% file: content/experiments.tex
\section{Experiments}
\label{sec:experiment}
We conduct a comprehensive evaluation of \method. The evaluation is divided into two main categories: speech tokenizer and speech generation. 
\subsection{Evaluation of Speech Tokenizer}

\subsubsection{Qwen-TTS-Tokenizer-25Hz}
As shown in Table \ref{tab:v1-asr}, we compare S3 Tokenizer series~\citep{cosyvoice,cosyvoice2}, also supervised semantic speech tokenizers, on automatic speech recognition (ASR) tasks across English and Chinese subsets of the CommonVoice (C.V.) and Fleurs benchmark. The Qwen-TTS-Tokenizer-25Hz variants demonstrate competitive performance. 
Qwen-TTS-Tokenizer-25Hz in the S1 stage (trained with ASR supervision) achieves ASR performance comparable to or better than the S3 Tokenizer series, attaining the lowest or near-lowest WER across multiple datasets. In the S2 stage, where the model is further fine-tuned to enhance the acoustic expressiveness of the tokens, ASR performance slightly degrades—consistent with expectations. This mild drop in recognition accuracy is attributed to the incorporation of additional acoustic details into the tokens, which, while reducing pure semantic discriminability, benefits downstream speech generation tasks such as high-quality TTS or waveform reconstruction, reflecting a deliberate trade-off between semantic fidelity and acoustic richness.
\begin{table}[H]
\setlength\tabcolsep{3pt}
\centering


\caption{\textbf{Comparison between different supervised semantic speech tokenizers on ASR Task. The highest scores are shown in bold. }}
\begin{tabular}{@{}lcccccc@{}}
\toprule
\textbf{Model}   & \multicolumn{1}{c}{\textbf{Codebook Size}} & \multicolumn{1}{c}{\textbf{FPS}} & \multicolumn{1}{c}{\textbf{C.V. EN}} & \multicolumn{1}{c}{\textbf{C.V. CN}} & \multicolumn{1}{c}{\textbf{Fluers EN}} & \multicolumn{1}{c}{\textbf{Fluers CN}}\\ \midrule
S3 Tokenizer(VQ)~\citep{cosyvoice}            &4096   &50   &12.06           & 15.38          & -             & -                                      \\
S3 Tokenizer(VQ)~\citep{cosyvoice}            &4096   &25   &11.56           & 18.26          & 7.65          & 5.03                                           \\
S3 Tokenizer(FSQ)~\citep{cosyvoice}           &6561   &25   &10.67           & \textbf{7.29}           & 6.58          & 4.43                                 \\
Qwen-TTS-Tokenizer-25Hz (Stage 1)   &32768  &25   &\textbf{7.51}            & 10.73          & \textbf{3.07}          & \textbf{4.23}                                    \\
Qwen-TTS-Tokenizer-25Hz (Stage 2)   &32768  &25   &10.40           & 14.99          & 4.14          & 4.67                               \\ \bottomrule

\end{tabular}
\label{tab:v1-asr}
\end{table}

\subsubsection{Qwen-TTS-Tokenizer-12Hz}
We evaluate speech reconstruction performance on the LibriSpeech test-clean set, which comprises 2,620 utterances. To enable fair comparison across models, we report key configuration parameters, including the number of quantizers (NQ), the codebook size, and the frame per second (FPS). Acoustic quality is assessed using Short-Time Objective Intelligibility (STOI), Perceptual Evaluation of Speech Quality (PESQ), and UTMOS, while speaker similarity (SIM) is measured with a WavLM-based speaker verification model. We compare Qwen-TTS-Tokenizer-12Hz against prior semantic-aware methods, including SpeechTokenizer~\citep{SpeechTokenizer}, XCodec series~\citep{xcodec,llasa}, XY-Tokenizer~\citep{xy-tokenzier}, Mimi~\citep{moshi}, and FireredTTS 2~\citep{FireRedTTS2}.
As shown in Table \ref{tab:v2-reconstruct}, Qwen-TTS-Tokenizer-12Hz not only sets a new state-of-the-art in speech reconstruction across all key metrics but, crucially, does so with remarkable encoding efficiency. This dual breakthrough in both quality and efficiency underscores the advanced capabilities of our method in speech representation learning and semantic information fusion.

\begin{table}[H]
\setlength\tabcolsep{2pt}
\footnotesize
\centering
\caption{\textbf{Comparison between different semantic-related speech tokenizers. The highest scores are shown in bold.}}
\begin{tabular}{@{}lcccccccc@{}}
\toprule
\textbf{Model}   & \multicolumn{1}{c}{\textbf{NQ}} & \multicolumn{1}{c}{\textbf{Codebook Size}} & \multicolumn{1}{c}{\textbf{FPS}} & \multicolumn{1}{c}{\textbf{PESQ\_WB}} & \multicolumn{1}{c}{\textbf{PESQ\_NB}} & \multicolumn{1}{c}{\textbf{STOI}} & \multicolumn{1}{c}{\textbf{UTMOS}} & \multicolumn{1}{c}{\textbf{SIM}}\\ \midrule
SpeechTokenizer~\citep{SpeechTokenizer}         & 8  &1024  &50   &2.60            & 3.05          & 0.92          & 3.90                    & 0.85                  \\
X-codec~\citep{xcodec}                 & 2  &1024  &50   &2.68            & 3.27          & 0.86          & 4.11                    & 0.84                          \\
X-codec 2~\citep{llasa}               & 1  &65536 &50   &2.43            & 3.04          & 0.92          & 4.13                    & 0.82                 \\
XY-Tokenizer~\citep{xy-tokenzier}            & 8  &1024  &12.5 &2.41            & 3.00          & 0.91          & 3.98                    & 0.83              \\
Mimi~\citep{moshi}                    & 16 &2048  &12.5 &2.88            & 3.42          & 0.94          & 3.87                    & 0.87                 \\
FireredTTS 2 Tokenizer~\citep{FireRedTTS2}             & 16 &2048  &12.5 &2.73            & 3.28          & 0.94          & 3.88                    & 0.87                    \\
Qwen-TTS-Tokenizer-12Hz   & 16 &2048  &12.5 &\textbf{3.21}   &\textbf{3.68}  &\textbf{0.96}  &\textbf{4.16}            & \textbf{0.95 }                           \\ \bottomrule

\end{tabular}
\label{tab:v2-reconstruct}
\end{table}

\subsection{Speech Generation}
In this section, we conduct a comprehensive evaluation of the speech generation capabilities of \method. To ensure a robust assessment across diverse scenarios, we categorize our experiments as follows:
\begin{itemize}
\item \textbf{Zero-Shot Speech Generation}: We evaluate the model's ability to clone unseen voices by measuring content consistency—specifically Word Error Rate (WER)—on the public Seed-TTS test set~\citep{seedtts}.
\item \textbf{Multilingual Speech Generation}: To assess linguistic versatility, we examine both content intelligibility and speaker similarity in a zero-shot multilingual setting using the multilingual test set from~\citep{minimax-speech}.
\item \textbf{Cross-Lingual Speech Generation}: We investigate the model's capacity for cross-lingual voice transfer (e.g., preserving timbre across language barriers) by evaluating content consistency on the CV3-Eval benchmark~\citep{cosyvoice3}.
\item \textbf{Controllable Speech Generation}: We verify the effectiveness of models' instruction-following on the InstructTTSEval benchmark~\citep{instructtts-eval}.
\item \textbf{Target-Speaker Speech Generation}: We analyze the generalization performance of our speaker fine-tuned (SFT) model variants on the multilingual test set~\citep{minimax-speech}, focusing on specific speaker adaptation.
\item \textbf{Long Speech Generation}: To validate the robustness and stability of our autoregressive architecture, we evaluate content consistency on an internal dataset consisting of generated speech samples exceeding 10 minutes in duration.
\end{itemize}

\subsubsection{Evaluation of Zero-Shot Speech Generation}
We conduct a comparative analysis of \method against leading state-of-the-art zero-shot TTS systems. Table~\ref{tab:zero_shot_speech_generation_table} reports the Word Error Rate (WER) as the primary metric for content consistency. The results highlight several key findings. 1): \method delivers robust performance across languages, attributed to the diverse acoustic data seen during pretraining and continual pretraining. 2): We observe that the 12Hz variants consistently outperform the 25Hz counterparts in terms of content accuracy (WER). This suggests that the coarser temporal resolution of the Qwen-TTS-Tokenizer-12Hz allows the autoregressive model to better model long-term dependencies for stable speech generation.3): Scaling the model size from 0.6B to 1.7B yields consistent gains. Specifically, after post-training, the \textbf{\method-12Hz-1.7B} variant achieves state-of-the-art performance on the \textit{test-en} set with a WER of 1.24, surpassing strong baselines like CosyVoice 3 and Seed-TTS.

\begin{table}[H]
\centering
\caption{\textbf{Zero-shot speech generation on the Seed-TTS test set. Performance is measured by Word Error Rate (WER, $\downarrow$), where lower is better. The best results are highlighted in bold.}}
\begin{tabular}{@{}cll@{}}
\toprule
\textbf{Datasets} & \textbf{Model} & \textbf{Performance} \\
\midrule
\multicolumn{3}{c}{\textit{Content Consistency}} \\
\midrule 
\multirow{14}{*}{\begin{tabular}[c]{@{}c@{}}\textbf{SEED} \\ \textit{test-zh} | \textit{test-en} \end{tabular}}   
   & Seed-TTS~\citep{seedtts} & 1.12 | 2.25  \\ 
   & MaskGCT~\citep{maskgct}                     & 2.27 | 2.62  \\ 
   & E2 TTS~\citep{e2tts}                        & 1.97 | 2.19  \\ 
   & F5-TTS~\citep{f5tts}                        & 1.56 | 1.83  \\ 
   & Spark TTS~\citep{sparktts}                  & 1.20 | 1.98  \\
   & Llasa-8B~\citep{llasa}                      & 1.59 | 2.97  \\
   & KALL-E~\citep{kalle}                      & 0.96 | 1.94  \\
   & FireRedTTS 2~\citep{FireRedTTS2}            & 1.14 | 1.95  \\
   & CosyVoice 3~\citep{cosyvoice3}            & \textbf{0.71} | 1.45  \\
   & MiniMax-Speech~\citep{minimax-speech}            & 0.83 | 1.65  \\
   & \method-25Hz-0.6B-Base                 & 1.18 | 1.64 \\
   & \method-25Hz-1.7B-Base                 & 1.10 | 1.49 \\
   & \method-12Hz-0.6B-Base                 & 0.92 | 1.32 \\
   & \method-12Hz-1.7B-Base                 & 0.77 | \textbf{1.24} \\
\bottomrule
\end{tabular}
\label{tab:zero_shot_speech_generation_table}
\end{table}

\subsubsection{Evaluation of Multilingual Speech Generation}
\method supports high-fidelity speech generation across 10 distinct languages. We benchmark its performance against leading commercial baselines, specifically MiniMax-Speech and ElevenLabs Multilingual v2. As detailed in Table~\ref{tab:multilingual_speech_generation_table}, \method achieves superior intelligibility (lowest WER) in 6 out of 10 languages, including Chinese, English, Italian, French, Korean, and Russian, surpassing the baselines by a significant margin. For the remaining languages (German, Portuguese, Spanish, and Japanese), \method maintains highly competitive performance comparable to the state-of-the-art. Furthermore, \method demonstrates dominant performance in voice cloning fidelity. It achieves the highest speaker similarity scores across all 10 evaluated languages, consistently outperforming both MiniMax-Speech and ElevenLabs. This superiority indicates that \method excels at capturing intrinsic speaker characteristics—such as timbre and prosody—while maintaining robust multilingual content generation.

\begin{table}[H]
\centering
\caption{\textbf{Multilingual speech generation on the TTS multilingual test set. Performance is measured by Word Error Rate (WER, $\downarrow$) for consistency and Cosine Similarity (SIM, $\uparrow$) for speaker similarity. The best results are highlighted in bold.}}
\begin{tabular}{@{} lcccccc @{}}
\toprule
\multirow{2}{*}{\textbf{Language}} & \multicolumn{2}{c}{\textbf{Qwen3-TTS-25Hz}} & \multicolumn{2}{c}{\textbf{Qwen3-TTS-12Hz}} & \multirow{2}{*}{\textbf{MiniMax}} & \multirow{2}{*}{\textbf{ElevenLabs}} \\
\cmidrule(lr){2-3} \cmidrule(lr){4-5}
& \textbf{0.6B-Base} & \textbf{1.7B-Base} & \textbf{0.6B-Base} & \textbf{1.7B-Base} & & \\ 
\midrule
\multicolumn{7}{c}{\textit{Content Consistency}} \\
\midrule
Chinese & 1.108 & \textbf{0.777} & 1.145 & 0.928 & 2.252 & 16.026 \\
English & 1.048 & 1.014 & \textbf{0.836} & 0.934 & 2.164 & 2.339 \\
German & 1.501 & 0.960 & 1.089 & 1.235 & 1.906 & \textbf{0.572} \\
Italian & 1.169 & 1.105 & 1.534 & \textbf{0.948} & 1.543 & 1.743 \\
Portuguese & 2.046 & 1.778 & 2.254 & 1.526 & 1.877 & \textbf{1.331} \\
Spanish & 2.031 & 1.491 & 1.491 & 1.126 & \textbf{1.029} & 1.084 \\
Japanese & 4.189 & 5.121 & 6.404 & 3.823 & \textbf{3.519} & 10.646 \\
Korean & 2.458 & 2.695 & \textbf{1.741} & 1.755 & 1.747 & 1.865 \\
French & 2.852 & \textbf{2.631} & 2.931 & 2.858 & 4.099 & 5.216 \\
Russian & 5.957 & 4.535 & 4.458 & \textbf{3.212} & 4.281 & 3.878 \\
\midrule
\multicolumn{7}{c}{\textit{Speaker Similarity}} \\
\midrule
Chinese & 0.797 & 0.796 & \textbf{0.811} & 0.799 & 0.780 & 0.677 \\
English & 0.811 & 0.815 & \textbf{0.829} & 0.775 & 0.756 & 0.613 \\
German & 0.749 & 0.737 & 0.769 & \textbf{0.775} & 0.733 & 0.614 \\
Italian & 0.722 & 0.718 & 0.792 & \textbf{0.817} & 0.699 & 0.579 \\
Portuguese & 0.790 & 0.783 & 0.794 & \textbf{0.817} & 0.805 & 0.711 \\
Spanish & 0.732 & 0.731 & 0.812 & \textbf{0.814} & 0.762 & 0.615 \\
Japanese & \textbf{0.810} & 0.807 & 0.798 & 0.788 & 0.776 & 0.738 \\
Korean & \textbf{0.824} & 0.814 & 0.812 & 0.799 & 0.776 & 0.700 \\
French & 0.698 & 0.703 & 0.700 & \textbf{0.714} & 0.628 & 0.535 \\
Russian & 0.734 & 0.744 & 0.781 & \textbf{0.792} & 0.761 & 0.676 \\
\bottomrule
\end{tabular}
\label{tab:multilingual_speech_generation_table}
\end{table}

\subsubsection{Evaluation of Cross-Lingual Speech Generation}
We assess the capability of \method to preserve speaker identity across language barriers (cross-lingual voice cloning). We benchmark against the CosyVoice series. Table~\ref{tab:cross_lingual_speech_generation_table} reports the error rates (WER/CER) for various source-target pairs. As shown in the table, \method establishes a new state-of-the-art in scenarios targeting English and Korean. Most notably, in \textit{zh-to-ko} generation, \method reduces the error rate by approximately 66\% compared to CosyVoice3 (4.82 vs. 14.4), demonstrating exceptional cross-lingual generalization. Besides, in frequently used translation pairs like \textit{zh-to-en} and \textit{en-to-zh}, \method outperforms baselines, indicating superior content consistency and reduced accent drift. While CosyVoice2 shows instability in several pairs, \method maintains consistently low error rates across all evaluated directions, confirming the robustness of our training strategy.

\begin{table}[H]
\centering
\caption{\textbf{Cross-lingual speech generation on the Cross-Lingual benchmark. Performance is measured by Mixed Error Rate (WER for English, CER for others, $\downarrow$). The best results are highlighted in bold.}}
\begin{tabular}{@{} lcccc @{}}
\toprule
\textbf{Task} & \textbf{\method-25Hz-1.7B-Base} & \textbf{\method-12Hz-1.7B-Base} & \textbf{CosyVoice3} & \textbf{CosyVoice2} \\
\midrule
en-to-zh & 5.66 & \textbf{4.77} & 5.09 & 13.5 \\
ja-to-zh & 3.92 & 3.43 & \textbf{3.05} & 48.1 \\
ko-to-zh & 1.14 & 1.08 & \textbf{1.06} & 7.70 \\
\midrule 
zh-to-en & 2.91 & \textbf{2.77} & 2.98 & 6.47 \\
ja-to-en & 3.95 & \textbf{3.04} & 4.20 & 17.1 \\
ko-to-en & 3.48 & \textbf{3.09} & 4.19 & 11.2 \\
\midrule
zh-to-ja & 9.29 & 8.40 & \textbf{7.08} & 13.1 \\
en-to-ja & 7.74 & 7.21 & \textbf{6.80} & 14.9 \\
ko-to-ja & 4.17 & \textbf{3.67} & 3.93 & 5.86 \\
\midrule
zh-to-ko & 8.12 & \textbf{4.82} & 14.4 & 24.8 \\
en-to-ko & 6.83 & \textbf{5.14} & 5.87 & 21.9 \\
ja-to-ko & 6.86 & \textbf{5.59} & 7.92 & 21.5 \\
\bottomrule
\end{tabular}
\label{tab:cross_lingual_speech_generation_table}
\end{table}

\subsubsection{Evaluation of Controllable Speech Generation}
We assess the instruction-following capabilities of \method using the InstructTTSEval benchmark. By adopting the ChatML format, \method treats voice control as a language modeling task, allowing for nuanced manipulation of speech attributes. The evaluation covers two distinct scenarios:
\textbf{Voice Design (Creation)}: In this scenario, the model generates novel voices based on text descriptions. As shown in Table~\ref{tab:controllable_speech_generation_table}, \textbf{\method-12Hz-1.7B-VD} establishes a new state-of-the-art among open-source models. Notably, it outperforms commercial systems like Hume and specialized models like VoiceSculptor in Description-Speech Consistency (DSD) and Response Precision (RP). This indicates superior alignment between the semantic input and the acoustic output.
\textbf{Target Speaker (Editing)}: This scenario tests the ability to modify attributes of a reference speaker. \method demonstrates robust performance, significantly outperforming GPT-4o-mini-tts across all metrics (e.g., +28\% APS improvement in Chinese). While the Gemini series remains a strong upper bound, \method exhibits competitive capability in preserving speaker identity while adhering to style modification instructions.

\begin{table}[H]
\setlength\tabcolsep{4pt}
\footnotesize
\centering
\caption{\textbf{Controllable speech generation on InstructTTSEval. Performance is measured by Attribute Perception and Synthesis accuracy (APS), Description–Speech Consistency (DSD), and Response Precision (RP). The highest scores are shown in bold.}}
\begin{tabular}{@{}llllllll@{}}
\toprule
\textbf{Type}                                                                      & \textbf{Model}                      & \multicolumn{3}{c}{\textbf{InstructTTSEval-ZH}}        & \multicolumn{3}{c}{\textbf{InstructTTSEval-EN}}        \\ \midrule
                                                                          &                            & APS (↑)       & DSD (↑)       & RP (↑)        & APS (↑)       & DSD (↑)       & RP (↑)        \\ \midrule
\multirow{4}{*}{\begin{tabular}[c]{@{}l@{}}\textit{Target}\\ \textit{Speaker}\end{tabular}} & Gemini-flash               & 88.2          & \textbf{90.9}          & \textbf{77.3}          & \textbf{92.3}          & \textbf{93.8}          & \textbf{80.1}          \\
                                                                          & Gemini-pro                 & \textbf{89.0}          & 90.1          & 75.5          & 87.6          & 86.0          & 67.2          \\
                                                                          & Qwen3TTS-25Hz-1.7B-CustomVoice      &  83.1             & 75.0              & 63.0              & 79.0              & 82.8              & 69.3     \\
                                                                          & Qwen3TTS-12Hz-1.7B-CustomVoice      &  83.0             & 77.8              & 61.2              & 77.3              & 77.1              & 63.7     \\
                                                                          & GPT-4o-mini-tts            & 54.9          & 52.3          & 46.0          & 76.4          & 74.3          & 54.8          \\
\midrule 
\multirow{9}{*}{\begin{tabular}[c]{@{}l@{}}\textit{Voice}\\ \textit{Design}\end{tabular}}   & Qwen3TTS-12Hz-1.7B-VD & \textbf{85.2} & \textbf{81.1} & \textbf{65.1} & 82.9 & \textbf{82.4} & \textbf{68.4} \\
                                                                          & Mimo-Audio-7B-Instruct~\citep{mimoaudio}     & 75.7          & 74.3          & 61.5          & 80.6          & 77.6          & 59.5          \\
                                                                          & VoiceSculptor~\citep{voicesculptor}              & 75.7          & 64.7          & 61.5          & -             & -             & -             \\
                                                                          & Hume                       & -             & -             & -             & \textbf{83.0}          & 75.3          & 54.3          \\
                                                                          & VoxInstruct~\citep{voxinstruct}                & 47.5          & 52.3          & 42.6          & 54.9          & 57.0          & 39.3          \\
                                                                          & Parler-tts-mini~\citep{parlertts}            & -             & -             & -             & 63.4          & 48.7          & 28.6          \\
                                                                          & Parler-tts-large~\citep{parlertts}           & -             & -             & -             & 60.0          & 45.9          & 31.2          \\
                                                                          & PromptTTS~\citep{prompttts}                  & -             & -             & -             & 64.3          & 47.2          & 31.4          \\
                                                                          & PromptStyle~\citep{promptstyle}                & -             & -             & -             & 57.4          & 46.4          & 30.9          \\ \bottomrule
\end{tabular}
\label{tab:controllable_speech_generation_table}
\end{table}

\subsubsection{Evaluation of Target-Speaker Speech Generation}
We assess the effectiveness of speaker fine-tuning for high-fidelity speaker adaptation. We fine-tune \method on a specific target speaker (Aiden Voice) and benchmark it against the GPT-4o-Audio-Preview (Ballad Voice) on the multilingual test set. As presented in Table~\ref{tab:single_speech_generation_table}, despite being fine-tuned exclusively on monolingual data, \method exhibits exceptional cross-lingual generalization. It successfully transfers the target speaker's timbre and prosody to all 10 evaluated languages without degradation in stability. Additionally, \method outperforms GPT-4o-Audio-Preview in 7 out of 10 languages. Specifically, it achieves lower Word Error Rates (WER) in Chinese, English, German, Spanish, Japanese, Korean, and Russian. While GPT-4o maintains a slight edge in Italian, Portuguese, and French, \method demonstrates remarkably better intelligibility in challenging languages like Japanese (3.88 vs. 5.00) and Korean (1.74 vs. 2.76), highlighting the robustness of our speaker fine-tuning strategy.

\begin{table}[H]
\footnotesize
\setlength\tabcolsep{4pt}
\centering
\caption{\textbf{Target-Speaker Multilingual Speech Generation on the TTS multilingual test set. Performance is measured by Word Error Rate (WER, $\downarrow$). The best results are highlighted in bold.}}
\label{tab:single_speech_generation_table}
\begin{tabular}{@{} lccccc @{}}
\toprule
\multirow{2}{*}{\textbf{Language}} & \multicolumn{2}{c}{\textbf{Qwen3-TTS-25Hz}} & \multicolumn{2}{c}{\textbf{Qwen3-TTS-12Hz}} & \multirow{2}{*}{\textbf{\makecell{GPT-4o-Audio\\Preview}}} \\ 
\cmidrule(lr){2-3} \cmidrule(lr){4-5}
 & \textbf{0.6B-CustomVoice} & \textbf{1.7B-CustomVoice} & \textbf{0.6B-CustomVoice} & \textbf{1.7B-CustomVoice} & \\ \midrule
Chinese    & 0.874 & \textbf{0.708} & 0.944 & 0.903 & 3.519 \\
English    & 1.332 & 0.936 & 1.188 & \textbf{0.899} & 2.197 \\
German     & 0.990 & \textbf{0.634} & 2.722 & 1.057 & 1.161 \\
Italian    & 1.861 & 1.271 & 2.545 & 1.362 & \textbf{1.194} \\
Portuguese & 1.728 & 1.854 & 3.219 & 2.681 & \textbf{1.504} \\
Spanish    & 1.309 & 1.284 & \textbf{1.154} & 1.330 & 4.000 \\
Japanese   & \textbf{3.875} & 4.518 & 6.877 & 4.924 & 5.001 \\
Korean     & 2.202 & 2.274 & 3.053 & \textbf{1.741} & 2.763 \\
French     & 3.865 & \textbf{3.080} & 3.841 & 3.781 & 3.605 \\
Russian    & 6.529 & \textbf{4.444} & 5.809 & 4.734 & 5.250 \\ \bottomrule
\end{tabular}
\end{table}

\subsubsection{Evaluation of Long Speech Generation}
Long-form synthesis presents unique challenges, often prone to issues like repetition, omission, or prosodic discontinuity. We evaluate this capability on a curated internal dataset comprising 100 texts in both Chinese and English, with lengths varying from 200 to 2000 words. Following the methodology of Seed-TTS Eval~\citep{seedtts}, we use Qwen3-ASR for transcription due to its high accuracy in long-form recognition. We compare the fine-tuned \method (Aiden Voice) against open-source baselines, including Higgs-Audio-v2, VibeVoice, and VoxCPM.
As detailed in Table~\ref{tab:long_speech_generation_table}, \method-25Hz-1.7B achieves the lowest WER across both languages (1.533 for \textit{long-zh} and 1.571 for \textit{long-en}). This demonstrates remarkable consistency compared to VibeVoice, which exhibits significant degradation in Chinese generation (WER > 22). Unlike chunk-based systems such as Higgs-Audio-v2 that suffer from boundary artifacts, \method generates seamless audio with consistent prosody throughout the entire duration. Notably, the 25Hz variant outperforms the 12Hz variant in this task, suggesting that semantic tokens may be more beneficial for maintaining stability over extended sequences.

\begin{table}[H]
\centering
\caption{\textbf{Long speech generation results. Performance is measured by Word Error Rate (WER, $\downarrow$). The best results are highlighted in bold.}}
\begin{tabular}{@{}cll@{}}
\toprule
\textbf{Datasets} & \textbf{Model} & \textbf{Performance} \\
\midrule
\multicolumn{3}{c}{\textit{Content Consistency}} \\
\midrule 
\multirow{5}{*}{\begin{tabular}[c]{@{}c@{}}\textit{long-zh}|\textit{long-en}\end{tabular}}   
   & Higgs-Audio-v2 (chunk)~\citep{higgsaudio2025}                    & 5.505 | 6.917  \\ 
   & VibeVoice~\citep{vibevoice}      & 22.619 | 1.780  \\ 
   & VoxCPM~\citep{voxcpm}                      & 4.835 | 7.474  \\ 
   & \method-25Hz-1.7B-CustomVoice                 & \textbf{1.517} | \textbf{1.225} \\
   & \method-12Hz-1.7B-CustomVoice                 & 2.356 | 2.812 \\
\bottomrule
\end{tabular}
\label{tab:long_speech_generation_table}
\end{table}

%% file: content/conclusion.tex
\section{Conclusion}
\label{sec:conclusion}
In this report, we introduced \method, a family of large-scale, multilingual, and robust text-to-speech models designed for real-time speech synthesis. Through a novel dual-track design two types of speech tokenizer comprising the semantic-rich \textit{Qwen-TTS-Tokenizer-25Hz} and the low-latency \textit{Qwen-TTS-Tokenizer-12Hz}, \method effectively synthesizes high-fidelity speech with streaming efficiency. Extensive evaluations confirm that our models achieve state-of-the-art performance across a wide spectrum of tasks. Specifically, \method sets new benchmarks in zero-shot voice cloning and cross-lingual synthesis, significantly outperforming existing baselines in challenging scenarios like Chinese-to-Korean generation. Besides, with the probabilistically activated \textit{thinking pattern}, \method sets new state-of-the-art performance in the voice design scenario. Furthermore, our dedicated training strategy resolves stability issues in autoregressive models, enabling the seamless generation of over 10 minutes of fluent speech without the artifacts typical of chunk-based systems.

\method unifies diverse speech generation tasks—ranging from zero-shot cloning and cross-lingual transfer to fine-grained instruction control—within a single autoregressive framework. This unification paves the way for the next generation of omni-capable audio systems. In future work, we aim to extend this architecture to support versatile audio generation, further scale our multilingual coverage beyond the current 10 languages, and explore more granular stylistic controls. By open-sourcing both the models and tokenizers, we hope to accelerate community research and facilitate the development of more natural, expressive, and accessible human-computer interfaces.

%% file: content/authors.tex
\section{Authors}
\textbf{Core Contributors:} Hangrui Hu, Xinfa Zhu, Ting He, Dake Guo, Bin Zhang, Xiong Wang, Zhifang Guo, Ziyue Jiang, Hongkun Hao, Zishan Guo, Xinyu Zhang, Pei Zhang, Baosong Yang, Jin Xu$^{\dag}$, Jingren Zhou, Junyang Lin$^{\dag}$

\textbf{Contributors\footnote{Alphabetical order. $^{\dag}$Corresponding Authors.}} Yunfei Chu, Daren Chen, Jiayi Leng, Zheng Li, Yuanjun Lv, Linhan Ma, Ziyang Ma, Xian Shi, Hao Su, Xuechun Wang, Yongqi Wang, Yuezhang Wang, Yuxuan Wang, Zhenglin Wang, Lei Xie, Kangxiang Xia, Qize Yang, Xian Yang, Jianwei Zhang, Guangdong Zhou, Jialong Zuo